\documentclass[11pt,a4paper]{article}

\usepackage[utf8]{inputenc}
\usepackage[T1]{fontenc}
\usepackage{lmodern}
\usepackage{microtype}
\usepackage[margin=2.0cm]{geometry}
\usepackage{setspace}
\usepackage{parskip}
\usepackage{graphicx}
\usepackage{booktabs}
\usepackage{enumitem}
\usepackage{hyperref}
\RequirePackage[]{xcolor}
\definecolor{AIPblue}      {RGB}{  0, 65,120}
\hypersetup{colorlinks=true,linkcolor=AIPblue,citecolor=AIPblue,urlcolor=AIPblue}
\usepackage{titling}
\usepackage{fancyhdr}
\usepackage{caption}
\usepackage{amsmath,amssymb}
\usepackage{natbib}
\usepackage[normalem]{ulem}
\usepackage{multicol}

\begin{document}

\thispagestyle{empty}

\begin{flushright}
{\bf\large White Paper \hfill \huge Expanding Horizons}

{\large\bf Transforming Astronomy in the 2040s}
\end{flushright}

\

Title:\vspace{-1.8em}\\
\vspace{-0.1em}\noindent\rule{\textwidth}{0.4pt}\\
{\LARGE \textbf{Large-scale time-series spectroscopy for stellar ages}\par}

\

Scientific Categories:\vspace{-1.0em}\\
\noindent\rule{\textwidth}{0.4pt}\\
{\large Spectroscopy, Stellar Physics, Time-domain, Stellar Evolution, Surveys\par}

\

Lead Authors:\vspace{-1.0em}\\
\noindent\rule{\textwidth}{0.4pt}
    {\bf David Gruner}     \hfill  Leibniz-Institute for Astrophysics Potsdam (AIP), Germany\\[-0.3em]
     \href{dgruner@aip.de}{dgruner[at]aip.de} \\
    \\[-0.5em]
    {\bf Sydney A. Barnes} \hfill Leibniz-Institute for Astrophysics Potsdam (AIP), Germany\\[-0.3em]
    \href{sbarnes@aip.de}{sbarnes[at]aip.de}\\
    \\[-0.5em]
    {\bf Ansgar Reiners}   \hfill Universit\"at G\"ottingen, Germany\\[-0.3em]
    \href{Ansgar.Reiners@phys.uni-goettingen.de}{Ansgar.Reiners[at]phys.uni-goettingen.de}

\

Contributing Authors:\vspace{-1.0em}\\
\noindent\rule{\textwidth}{0.4pt}\\ 
{\small\begin{tabular}{p{.6\linewidth}l}
    Klaus G. Strassmeier    \hfill \href{kstrassmeier@aip.de}{kstrassmeier[at]aip.de}                     & Leibniz-Inst. for Astrophys. Potsdam\\
    Cristina Chiappini      \hfill \href{cristina.chiappini@aip.de}{cristina.chiappini[at]aip.de}         & Leibniz-Inst. for Astrophys. Potsdam\\
    J\"org Weingrill        \hfill \href{jweingrill@aip.de}{jweingrill[at]aip.de}                         & Leibniz-Inst. for Astrophys. Potsdam\\
    Michael Weber           \hfill \href{mweber@aip.de}{mweber[at]aip.de}                                 & Leibniz-Inst. for Astrophys. Potsdam\\
    Ilya Ilyin              \hfill \href{ilyin@aip.de}{ilyin[at]aip.de}                                   & Leibniz-Inst. for Astrophys. Potsdam\\
    Thomas Granzer          \hfill \href{tgranzer@aip.de}{tgranzer[at]aip.de}                             & Leibniz-Inst. for Astrophys. Potsdam\\
    \"Ozg\"un Adebali       \hfill \href{oadebali@aip.de}{oadebali[at]aip.de}                             & Leibniz-Inst. for Astrophys. Potsdam\\
    Jean-Michel D\'esert    \hfill \href{jmdesert@aip.de}{jmdesert[at]aip.de}                             & Leibniz-Inst. for Astrophys. Potsdam\\
    Marica Valentini        \hfill \href{mvalentini@aip.de}{mvalentini[at]aip.de}                         & Leibniz-Inst. for Astrophys. Potsdam\\
    Dario Fritzewski        \hfill \href{dario.fritzewski@kuleuven.be}{dario.fritzewski[at]kuleuven.be}   & KU Leuven, Belgium\\
    Paolo Ventura           \hfill \href{paolo.ventura@inaf.it}{paolo.ventura[at]inaf.it}                 & INAF - Osserv. Astron. di Roma\\
    Alfio Bonanno           \hfill \href{alfio.bonanno@inaf.it}{alfio.bonanno[at]inaf.it}                 & INAF - Osserv. Astron. di Catania\\
    Jose-Dias do Nascimento \hfill \href{jdonascimento@cfa.harvard.edu}{jdonascimento[at]cfa.harvard.edu} & Harvard-Smithsonian CfA, USA\\
    Jorge Melendez          \hfill \href{jorge.melendez@iag.usp.br}{jorge.melendez[at]iag.usp.br}         & IAG, Univ. Sao Paolo, Brazil\\
    Santosh Joshi           \hfill \href{santosh@aries.res.in}{santosh[at]aries.res.in}                   & ARIES Observatory, Nainital, India \\
    Yong-Cheol Kim          \hfill \href{yckim@yonsei.ac.kr}{yckim[at]yonsei.ac.kr}                       & Yonsei Univ., Seoul, Korea\\
\end{tabular}}

\

\

\newpage

\setcounter{page}{1}

\newpage

\section*{Scientific challenges in the 2040s}

    To date, Galactic Astronomy has largely concerned itself with astrophysical processes, and with the locations, space motions and compositions of objects.
    Consider, for example, the elucidation of the components of the Galaxy over the past decades, its mapping as enabled by \textit{Gaia} and its predecessors, the photometric and spectroscopic characterization of innumerable astrophysical objects in various wavelength ranges, both from the ground and from space, and the expanding discovery and characterization of exoplanets; all focused on the current, static Galaxy.
    This White Paper proposes a dedicated program to derive stellar ages from time-series spectroscopy to hasten the transformation of this static conception into a dynamical one with age-labeled objects and events.

    While stars have been recognized for about 75 years to have evolved over their lives, and cluster stars and even certain individual stars have been dated using classical isochrones \citep{1953AJ.....58...61S,1964ApJ...140..544D}, it is only comparatively recently that it has become possible even to contemplate the possibility of dating individual main sequence field stars \citep[e.g. review by][]{2010ARAnA..48..581S}.
    An observational astronomical future is imminent where \emph{every star} in a catalogue is labeled with a \emph{measured quantity} such as a brightness/colour combination, frequency spacing, rotation period, lithium or other element abundance (or something else entirely) that enables its age to be estimated\footnote{Note that it is the relevant observed quantity that must be measured and provided; the age itself is inferred, and its determination, together with the precision of that determination, will necessarily evolve naturally in the literature.}. 
    Although other procedures will also contribute ages of varying reliability for specific types of stars, asteroseismology and gyrochronology have largely allowed the conceit of this recent possibility to be articulated and embodied \citep[e.g. in the \textsc{Plato} mission,][]{2014ExA....38..249R} because they work on \emph{main sequence stars}, a particularly difficult group of stars to derive ages for.   
    Such age determination will be revolutionary for Galactic Astronomy because it will enable the sequence of assemblage and unfolding of astrophysical processes to be explicated in detail, enabling the construction of histories.

    Asteroseismology is currently almost entirely based on data from space, and inconceivable for the desired/required numbers of stars. 
    The observational centre-of-gravity for rotation period measurements has also lurched towards space-based data, nominally optimized for other purposes such as exoplanet discovery.
    However, even space photometry cannot provide the necessary measurements for many targets, because for stars more mature or warmer than the Sun the photometric variability itself becomes negligible.
    The way forward \citep[as has been known since the Mt.\,Wilson program, e.g.,][]{1957ApJ...125..661W,1996ApJ...457L..99B}
    is to transition from photometry to spectroscopy.
    This White Paper articulates the case that ESO could play a vital role in constructing a \emph{``Chronology of the Galaxy''}, with a dedicated facility/program to obtain the requisite time-domain spectroscopy to then enable the derivation of rotational ages (and ancillary information\footnote{Over time, stellar cycles, and the patterns therein, will also become apparent, possibly on a massive scale, creating a legacy sample that will enable a deeper understanding of stellar magnetism, magnetohydrodynamics, and internal flows on stars, leading to an intrinsically dynamical understanding of the workings of stars.}) for such objects in the 2040s.

\section*{Relevant science themes for the ESO community and beyond }
    
    Creating {\bf age-labeled Galactic stellar populations} will have transformational effects on Galactic Astronomy.
    The coming decades will find us building upon the legacy of the \textit{Gaia} mission, mapping the component building blocks of the Galaxy, identifying remnants of past mergers and phases of star formation, fueling a revolution in Galactic Archaeology. 
    Once the compositions of these populations have been mapped out, for instance, with facilities such as 4MOST, it will be essential to age-date them for further progress. 

    A spectacular recent application of gyrochronology was the identification of the Greater Pleiades Complex, an over 3,000-member all-sky group of sister stars of the eponymous naked-eye open cluster, that used gyrochronology to identify the coevality of these young stars \citep{2025ApJ...994...24B}.
    Similar {\bf identifications of new age-distinct populations} in the Galaxy were obtained for Group X \citep{2022A&A...657L...3M} and the Pisces-Eridanus Stellar Stream \citep{2019AJ....158...77C} using gyrochronology.
    These identifications would not have been possible without rotation period measurements from TESS and Kepler satellite data. 
    This process of identification of coeval stellar siblings will accelerate in the next decades to older populations, \emph{but only if the requisite measurements are available.} 

    Once acquired, the large numbers of stars with associated ages will enable the investigation of numerous science themes, some of which are:

    \begin{description}

        \item[The origins, past, and future of the Sun:]
            Humanity will be tethered to the Sun for the foreseeable future, and as such, the Sun will remain a focal point for astronomical research.
            Extensive chronologies of sun-like stars will reveal the detailed history of the likely evolution of our Sun.
            Moreover, ages for sufficient numbers of stars will eventually lead to the identification of the Sun's siblings and, combined with the 6D-mapped Galaxy by e.g., \textit{Gaia}, the star cluster that it was likely born within.
            It is conceivable that this (evidently dispersed) cluster will be identified in the next generation, if not within the current one itself.

        \item[A chronology of exoplanet evolution:]
            Exoplanet research (and detection) has seen an incredible boom in the last decade.
            Our understanding has, however, not kept pace.
            Building our understanding of the nature and evolution of exoplanets will be a task for the coming decades.
            Understanding planet host stars is crucial to this endeavour, and knowledge of their ages will enable exoplanet chronologies. Those, in turn, will be necessary to reveal the intricate details of their evolution from formation to ultimate fate, and provide context for discoveries made by e.g., the HWO and LIFE missions.
            Furthermore, inferring an exoplanet's age via its host star will be essential if/when signs of life are detected outside the solar system.
            Time series spectroscopy also opens up additional possibilities for quantitative studies of the impact of host star variability on e.g., an exoplanet's atmospheric evolution, physical properties, and climate.
            
        \item[Consistency in stellar ages: ] 
            Consistency in stellar ages across multiple methods will be essential to building confidence in the relevant methods.
            This will make it necessary to study, not just clusters but also large numbers of systems such as wide binaries, where the individual components allow multiple age determination methods to be applied simultaneously, thereby also enabling sensitive methodological tests. Time domain spectroscopy of such targets is key here.

        \item[Advances in stellar physics: ]            
            Stellar ages for diverse and extensive samples of stars can be assembled into chrononologies that illustrate temporal evolution of objects and processes despite the static picture of the Galaxy accessible to us mortals.
            Such chronologies will lead to significant advances in stellar physics, including in understanding rotation, activity, convection, magnetism, and the dynamical properties of stars.
            Other objects for which ages cannot directly be obtained, such as different types of (e.g. more massive) stars or sub-stellar objects that can still be kinematically associated with age-dated stars, could also then be placed in appropriate chronologies.
            Time series spectroscopy will also enable additional science such as robust chromospheric ages (using both averages, and future developments), identification of stellar cycles (including cycle periods if the time baseline is long enough), etc.

    \end{description}

\section*{The case for time-series spectroscopy}

    Gyrochronology leverages the stellar rotation period (together with a proxy for stellar mass, typically the color) to derive stellar ages.
    The required rotation periods are usually measured by monitoring periodic brightness fluctuations induced by star spots traversing the stellar disk.
    Numerous ground-based (e.g., NOAO, ASAS-SN, and KELT) and space-based (e.g., CoRoT, Kepler, \textit{Gaia}, TESS) surveys have provided hundreds of thousands of rotation periods, and \textsc{PLATO} will soon add further to that.
    With very few exceptions, however, these measurements are performed for only fast rotating (i.e. younger) stars.
    For older (i.e., slower rotating) stars the photometric modulation is so weak that even the most sensitive photometric instruments are unable to measure it.
    The Sun, about half-way through its main sequence lifetime, falls into this group.
    Thus, the central question of this white paper is:\\
    \textit{How can rotation periods -- and ages -- be obtained efficiently for large numbers of such stars?}

    The answer, as demonstrated by the Mount Wilson program \citep[e.g.][]{1957ApJ...125..661W, 2007ApJS..171..260L}, is via {\bf monitoring the Ca\,\textsc{II}\,H\,\&\,K chromospheric emission lines}, whose modulations can be measured to yield rotation periods (and additional variations such as magnetic cycles) even when spot modulation is too weak \citep[e.g.,][]{1996ApJ...457L..99B}.
    This white paper urges an efficient and optimized survey that vastly expands and improves upon the original program.
    The Mt.\,Wilson survey performed narrow band photometry (using unconventionally designed filters) to measure flux ratios between the stellar continuum and the Ca\,\textsc{ii} H \& K line cores over the course of decades for some hundreds of stars.
    Many details of the survey including those concerning calibration are unfortunately lost, making reproducibility of the data very difficult and reducing inter-comparability between observations.

    The main improvement we envision is simply to move from the narrow-band photometry in the original program to low-resolution spectroscopy. 
    This comes with certain advantages: (1) Emission core fluxes can always be re-calibrated against the rest of the spectrum, (2) spectroscopy can account for radial velocity shifts, and (3) problems with source confusion, and subsequent discoveries of the multiplicity of targets can be mitigated as the spectra obtained contain additional information about the target.
    Spectroscopic time series also allow additional aspects of the targets to be investigated, such as other activity indicators besides Ca\,\textsc{ii}\,H\,\&\,K (depending on spectral coverage), radial velocity shifts that reveal multiplicity, or the variation of other stellar parameters and spectral features (e.g., induced by activity cycles).
    The feasibility of this approach for deriving the rotation period and age of an older solar twin has been recently demonstrated \citep{2025ApJ...983L..31C} \emph{using ESO observations}, and our goal is to expand such work vastly.

    Existing facilities, such as 4MOST or LAMOST are not designed to operate in the time domain and have limited blue sensitivity.
    HARPS, at La Silla's 3.6\,m telescope, operates in the time domain, but its high spectral resolution is disadvantageous for the purposes envisioned here.

    Therefore, we suggest the development of a facility dedicated to spectroscopic monitoring of stars that is optimized for blue light (Ca\,\textsc{ii}\,H\&K is located at 395\,nm).
    Only low resolution ($R\leq5\,000$) spectroscopy is required, with the associated advantages.
    Such a facility would benefit greatly from automated/robotic operation.
    Multi-object spectroscopy is possible if blue sensitivity can be maintained (e.g., with specialized fibers) and would have the potential to multiply the survey's impact.
    In the (least desirable) case of a single, medium-sized telescope that observes on a star-by-star basis, such a survey could provide rotation periods for several thousand stars per year that are otherwise unobtainable.
    The location of such a facility is of secondary importance, although the pursuit of all-sky coverage (by using multiple facilities at different locations) is desirable, permitting the study of diverse stellar populations in the Galaxy to obtain the most complete picture.

\begin{multicols}{2}[]
 \setlength{\bibsep}{0.0em}
 \footnotesize\bibliographystyle{aa}
 \bibliography{refs} 

@ARTICLE{1957ApJ...125..661W,
       author = {{Wilson}, O.~C. and {Vainu Bappu}, M.~K.},
        title = "{H and K Emission in Late-Type Stars: Dependence of Line Width on Luminosity and Related Topics.}",
      journal = {\apj},
         year = 1957,
        month = may,
       volume = {125},
        pages = {661},
          doi = {10.1086/146339},
       adsurl = {https://ui.adsabs.harvard.edu/abs/1957ApJ...125..661W}
}

@ARTICLE{1964ApJ...140..544D,
       author = {{Demarque}, P.~R. and {Larson}, R.~B.},
        title = "{The Age of Galactic Cluster NGC 188.}",
      journal = {\apj},
         year = 1964,
        month = aug,
       volume = {140},
        pages = {544},
          doi = {10.1086/147948},
       adsurl = {https://ui.adsabs.harvard.edu/abs/1964ApJ...140..544D}
}

@ARTICLE{1953AJ.....58...61S,
       author = {{Sandage}, A.~R.},
        title = "{The color-magnitude diagram for the globular cluster M 3.}",
      journal = {\aj},
         year = 1953,
        month = jan,
       volume = {58},
        pages = {61-75},
          doi = {10.1086/106822},
       adsurl = {https://ui.adsabs.harvard.edu/abs/1953AJ.....58...61S}
}

@ARTICLE{1996ApJ...457L..99B,
   author = {{Baliunas}, S. and {Sokoloff}, D. and {Soon}, W.},
    title = "{Magnetic Field and Rotation in Lower Main-Sequence Stars: an Empirical Time-dependent Magnetic Bode's Relation?}",
  journal = {\apjl},
     year = 1996,
    month = feb,
   volume = 457,
    pages = {L99},
      doi = {10.1086/309891},
   adsurl = {https://ui.adsabs.harvard.edu/abs/1996ApJ...457L..99B}
}

@ARTICLE{2007ApJS..171..260L,
       author = {{Lockwood}, G.~W. and {Skiff}, B.~A. and {Henry}, Gregory W. and {Henry}, Stephen and {Radick}, R.~R. and {Baliunas}, S.~L. and {Donahue}, R.~A. and {Soon}, W.},
        title = "{Patterns of Photometric and Chromospheric Variation among Sun-like Stars: A 20 Year Perspective}",
      journal = {\apjs},
         year = 2007,
        month = jul,
       volume = {171},
       number = {1},
        pages = {260-303},
          doi = {10.1086/516752},
archivePrefix = {arXiv},
       eprint = {astro-ph/0703408},
 primaryClass = {astro-ph},
       adsurl = {https://ui.adsabs.harvard.edu/abs/2007ApJS..171..260L}
}

@ARTICLE{2010ARAnA..48..581S,
       author = {{Soderblom}, David R.},
        title = "{The Ages of Stars}",
      journal = {\araa},
         year = 2010,
        month = sep,
       volume = {48},
        pages = {581-629},
          doi = {10.1146/annurev-astro-081309-130806},
archivePrefix = {arXiv},
       eprint = {1003.6074},
 primaryClass = {astro-ph.SR},
       adsurl = {https://ui.adsabs.harvard.edu/abs/2010ARA&A..48..581S}
}

@ARTICLE{2025ApJ...983L..31C,
       author = {{Carvalho-Silva}, Gabriela and {Mel{\'e}ndez}, Jorge and {Rathsam}, Anne and {Shejeelammal}, J. and {Martos}, Giulia and {Lorenzo-Oliveira}, Diego and {Spina}, Lorenzo and {Ribeiro Alves}, D{\'e}bora},
        title = "{A New Age{\textendash}Activity Relation For Solar Analogs that Accounts for Metallicity}",
      journal = {\apjl},
         year = 2025,
        month = apr,
       volume = {983},
       number = {2},
          eid = {L31},
        pages = {L31},
          doi = {10.3847/2041-8213/adc382},
archivePrefix = {arXiv},
       eprint = {2504.17482},
 primaryClass = {astro-ph.SR},
       adsurl = {https://ui.adsabs.harvard.edu/abs/2025ApJ...983L..31C}
}

@ARTICLE{2014ExA....38..249R,
       author = {{Rauer}, H. and {Catala}, C. and {Aerts}, C. and {Appourchaux}, T. and
         {Benz}, W. and {Brandeker}, A. and {Christensen-Dalsgaard}, J. and
         {Deleuil}, M. and {Gizon}, L. and {Goupil}, M. -J. and {G{\"u}del}, M. and
         {Janot-Pacheco}, E. and {Mas-Hesse}, M. and {Pagano}, I. and
         {Piotto}, G. and {Pollacco}, D. and {Santos}, {\.{C}}. and {Smith}, A. and
         {Su{\'a}rez}, J. -C. and {Szab{\'o}}, R. and {Udry}, S. and
         {Adibekyan}, V. and {Alibert}, Y. and {Almenara}, J. -M. and
         {Amaro-Seoane}, P. and {Eiff}, M. Ammler-von and {Asplund}, M. and
         {Antonello}, E. and {Barnes}, S. and {Baudin}, F. and {Belkacem}, K. and
         {Bergemann}, M. and {Bihain}, G. and {Birch}, A.~C. and {Bonfils}, X. and
         {Boisse}, I. and {Bonomo}, A.~S. and {Borsa}, F. and {Brand
        {\~a}o}, I.~M. and {Brocato}, E. and {Brun}, S. and {Burleigh}, M. and
         {Burston}, R. and {Cabrera}, J. and {Cassisi}, S. and {Chaplin}, W. and
         {Charpinet}, S. and {Chiappini}, C. and {Church}, R.~P. and
         {Csizmadia}, Sz. and {Cunha}, M. and {Damasso}, M. and {Davies}, M.~B. and
         {Deeg}, H.~J. and {D{\'\i}az}, R.~F. and {Dreizler}, S. and
         {Dreyer}, C. and {Eggenberger}, P. and {Ehrenreich}, D. and
         {Eigm{\"u}ller}, P. and {Erikson}, A. and {Farmer}, R. and
         {Feltzing}, S. and {de Oliveira Fialho}, F. and {Figueira}, P. and
         {Forveille}, T. and {Fridlund}, M. and {Garc{\'\i}a}, R.~A. and
         {Giommi}, P. and {Giuffrida}, G. and {Godolt}, M. and
         {Gomes da Silva}, J. and {Granzer}, T. and {Grenfell}, J.~L. and
         {Grotsch-Noels}, A. and {G{\"u}nther}, E. and {Haswell}, C.~A. and
         {Hatzes}, A.~P. and {H{\'e}brard}, G. and {Hekker}, S. and
         {Helled}, R. and {Heng}, K. and {Jenkins}, J.~M. and {Johansen}, A. and
         {Khodachenko}, M.~L. and {Kislyakova}, K.~G. and {Kley}, W. and
         {Kolb}, U. and {Krivova}, N. and {Kupka}, F. and {Lammer}, H. and
         {Lanza}, A.~F. and {Lebreton}, Y. and {Magrin}, D. and
         {Marcos-Arenal}, P. and {Marrese}, P.~M. and {Marques}, J.~P. and
         {Martins}, J. and {Mathis}, S. and {Mathur}, S. and {Messina}, S. and
         {Miglio}, A. and {Montalban}, J. and {Montalto}, M. and
         {Monteiro}, M.~J.~P.~F.~G. and {Moradi}, H. and {Moravveji}, E. and
         {Mordasini}, C. and {Morel}, T. and {Mortier}, A. and {Nascimbeni}, V. and
         {Nelson}, R.~P. and {Nielsen}, M.~B. and {Noack}, L. and
         {Norton}, A.~J. and {Ofir}, A. and {Oshagh}, M. and {Ouazzani}, R. -M. and
         {P{\'a}pics}, P. and {Parro}, V.~C. and {Petit}, P. and {Plez}, B. and
         {Poretti}, E. and {Quirrenbach}, A. and {Ragazzoni}, R. and
         {Raimondo}, G. and {Rainer}, M. and {Reese}, D.~R. and {Redmer}, R. and
         {Reffert}, S. and {Rojas-Ayala}, B. and {Roxburgh}, I.~W. and
         {Salmon}, S. and {Santerne}, A. and {Schneider}, J. and {Schou}, J. and
         {Schuh}, S. and {Schunker}, H. and {Silva-Valio}, A. and
         {Silvotti}, R. and {Skillen}, I. and {Snellen}, I. and {Sohl}, F. and
         {Sousa}, S.~G. and {Sozzetti}, A. and {Stello}, D. and
         {Strassmeier}, K.~G. and {{\v{S}}vanda}, M. and {Szab{\'o}}, Gy. M. and
         {Tkachenko}, A. and {Valencia}, D. and {Van Grootel}, V. and
         {Vauclair}, S.~D. and {Ventura}, P. and {Wagner}, F.~W. and
         {Walton}, N.~A. and {Weingrill}, J. and {Werner}, S.~C. and
         {Wheatley}, P.~J. and {Zwintz}, K.},
        title = "{The PLATO 2.0 mission}",
      journal = {Experimental Astronomy},
         year = 2014,
        month = nov,
       volume = {38},
       number = {1-2},
        pages = {249-330},
          doi = {10.1007/s10686-014-9383-4},
archivePrefix = {arXiv},
       eprint = {1310.0696},
 primaryClass = {astro-ph.EP},
       adsurl = {https://ui.adsabs.harvard.edu/abs/2014ExA....38..249R}
}

@ARTICLE{2019AJ....158...77C,
       author = {{Curtis}, Jason L. and {Ag{\"u}eros}, Marcel A. and {Mamajek}, Eric E. and {Wright}, Jason T. and {Cummings}, Jeffrey D.},
        title = "{TESS Reveals that the Nearby Pisces-Eridanus Stellar Stream is only 120 Myr Old}",
      journal = {\aj},
         year = 2019,
        month = aug,
       volume = {158},
       number = {2},
          eid = {77},
        pages = {77},
          doi = {10.3847/1538-3881/ab2899},
archivePrefix = {arXiv},
       eprint = {1905.10588},
 primaryClass = {astro-ph.SR},
       adsurl = {https://ui.adsabs.harvard.edu/abs/2019AJ....158...77C}
}

@ARTICLE{2022A&A...657L...3M,
       author = {{Messina}, S. and {Nardiello}, D. and {Desidera}, S. and {Baratella}, M. and {Benatti}, S. and {Biazzo}, K. and {D'Orazi}, V.},
        title = "{Gyrochronological dating of the stellar moving group Group X}",
      journal = {\aap},
         year = 2022,
        month = jan,
       volume = {657},
          eid = {L3},
        pages = {L3},
          doi = {10.1051/0004-6361/202142276},
archivePrefix = {arXiv},
       eprint = {2112.08061},
 primaryClass = {astro-ph.SR},
       adsurl = {https://ui.adsabs.harvard.edu/abs/2022A&A...657L...3M}
}

@ARTICLE{2025ApJ...994...24B,
       author = {{Boyle}, Andrew W. and {Bouma}, Luke G. and {Mann}, Andrew W.},
        title = "{Lost Sisters Found: TESS and Gaia Reveal a Dissolving Pleiades Complex}",
      journal = {\apj},
         year = 2025,
        month = nov,
       volume = {994},
       number = {1},
          eid = {24},
        pages = {24},
          doi = {10.3847/1538-4357/ae0724},
archivePrefix = {arXiv},
       eprint = {2511.07533},
 primaryClass = {astro-ph.SR},
       adsurl = {https://ui.adsabs.harvard.edu/abs/2025ApJ...994...24B}
}
\end{multicols}

\end{document}